\documentclass[11pt]{amsart}
\usepackage[all]{xy}
\usepackage[dvips]{graphicx}
\usepackage{amsfonts}
\usepackage{amssymb,latexsym,amsmath}
\usepackage{amsthm}
\usepackage{color}
\usepackage{empheq}
\usepackage{float}
\usepackage{hyperref}
\usepackage{listings}
\usepackage{mathrsfs}
\usepackage{slashed}
\usepackage{tikz}

\definecolor{dkgreen}{rgb}{0,0.6,0}
\definecolor{gray}{rgb}{0.5,0.5,0.5}
\definecolor{mauve}{rgb}{0.58,0,0.82}

\theoremstyle{definition}
\newtheorem{remark}{Remark}

\begin{document}
\title{Cohort Revenue \& Retention Analysis: A Bayesian Approach}
\author{Juan Camilo Orduz}
\email{juanitorduz@gmail.com}
\urladdr{\href{https://juanitorduz.github.io/}{https://juanitorduz.github.io/}}
\address{Berlin, Germany}
\date{\today}

\begin{abstract}
    We present a Bayesian approach to model cohort-level retention rates and revenue over time. We use Bayesian additive
    regression trees (BART) to model the retention component which we couple with a linear model for the revenue component.
    This method is flexible enough to allow adding additional covariates to both model components. This Bayesian framework
    allows us to quantify uncertainty in the estimation, understand the effect of covariates on retention through partial
    dependence plots (PDP) and individual conditional expectation (ICE) plots, and most importantly, forecast future
    revenue and retention rates with well-calibrated uncertainty through highest density intervals. We also provide
    alternative approaches to model the retention component using neural networks and inference through stochastic variational
    inference.
\end{abstract}

\maketitle

\addtocontents{toc}{\protect\setcounter{tocdepth}{1}}

\section{Introduction}

Understanding and predicting customer behavior directly impacts business profitability through improved retention strategies
and resource allocation. Among the metrics that define business success, retention and customer lifetime value
estimation stand at the forefront, serving as critical indicators of a company's ability to not only attract but maintain a
loyal customer base. These metrics transcend mere financial accounting—they represent the foundation upon which long-term
business strategies are built and refined. Seminal work by Fader and Hardie has established frameworks for both contractual
settings \cite{FaderHardie2007}, where subscription-based relationships predominate, and non-contractual settings
\cite{FaderHardie2005}, where customers may come and go without formal notification\footnote{Our definition of retention
    corresponds to what they call survival curve. See precise definitions below.}. Modern implementations of these CLV
models can now be found in Bayesian probabilistic programming frameworks such as PyMC (\cite{pymc2023}), where the
PyMC-Marketing library \cite{pymc_marketing} provides implementations of many standard buy-till-you-die (BTYD) models including
the BG/NBD, Pareto/NBD, and Gamma-Gamma models in a flexible, Bayesian framework. While these approaches have proven very valuable, they
often struggle to scale effectively. They can definitively be scaled with modern hardware and algorithms (for example,
stochastic variational inference, as described below). Nevertheless, this requires non-trivial work and effort.\\

For many decision-making processes, companies just need to understand behaviors at the cohort level—groups of customers
who joined during the same time period. In this paper we focus on this level of granularity. When shifting from individual
to cohort-level analysis, businesses typically face a methodological trilemma:

\begin{enumerate}
    \item \textbf{Complete pooling}: Aggregate all cohorts together and model retention and revenue as a collective whole,
          potentially obscuring important cohort-specific patterns.

    \item \textbf{No pooling}: Analyze each cohort in isolation, potentially overlooking valuable cross-cohort information
          and suffering from data sparsity for newer cohorts.

    \item \textbf{Partial pooling}: Model cohorts jointly with shared parameters, striking a balance between cohort-specific
          insights and statistical power.
\end{enumerate}

As detailed by \cite{FaderHardieNote2017}, each approach offers distinct advantages and limitations. However, a fundamental
challenge persists across these traditional methodologies: they typically lack the flexibility to efficiently incorporate
seasonality patterns and external regressors\footnote{Although, one can add regressors in some cases as described in
    \cite{FaderHardieNote2007} for the non-contractual case.}. This limitation becomes particularly problematic for businesses
with highly seasonal customer behavior—from retail operations affected by holiday shopping patterns to subscription services
influenced by annual promotional cycles. While some might argue that seasonality is secondary when estimating customer
lifetime value, the reality for many business models is that seasonal fluctuations significantly impact customer acquisition,
engagement, and retention decisions. Beyond the methodological challenges, businesses face practical hurdles in translating
retention and revenue models into actionable insights. Static models that fail to adapt to changing market dynamics or
consumer preferences quickly become outdated. Moreover, point estimates without associated uncertainty measures can lead to
misplaced confidence in business forecasts, potentially resulting in suboptimal resource allocation and strategic planning. \\

\begin{figure}
    \centering
    \includegraphics[width=\textwidth]{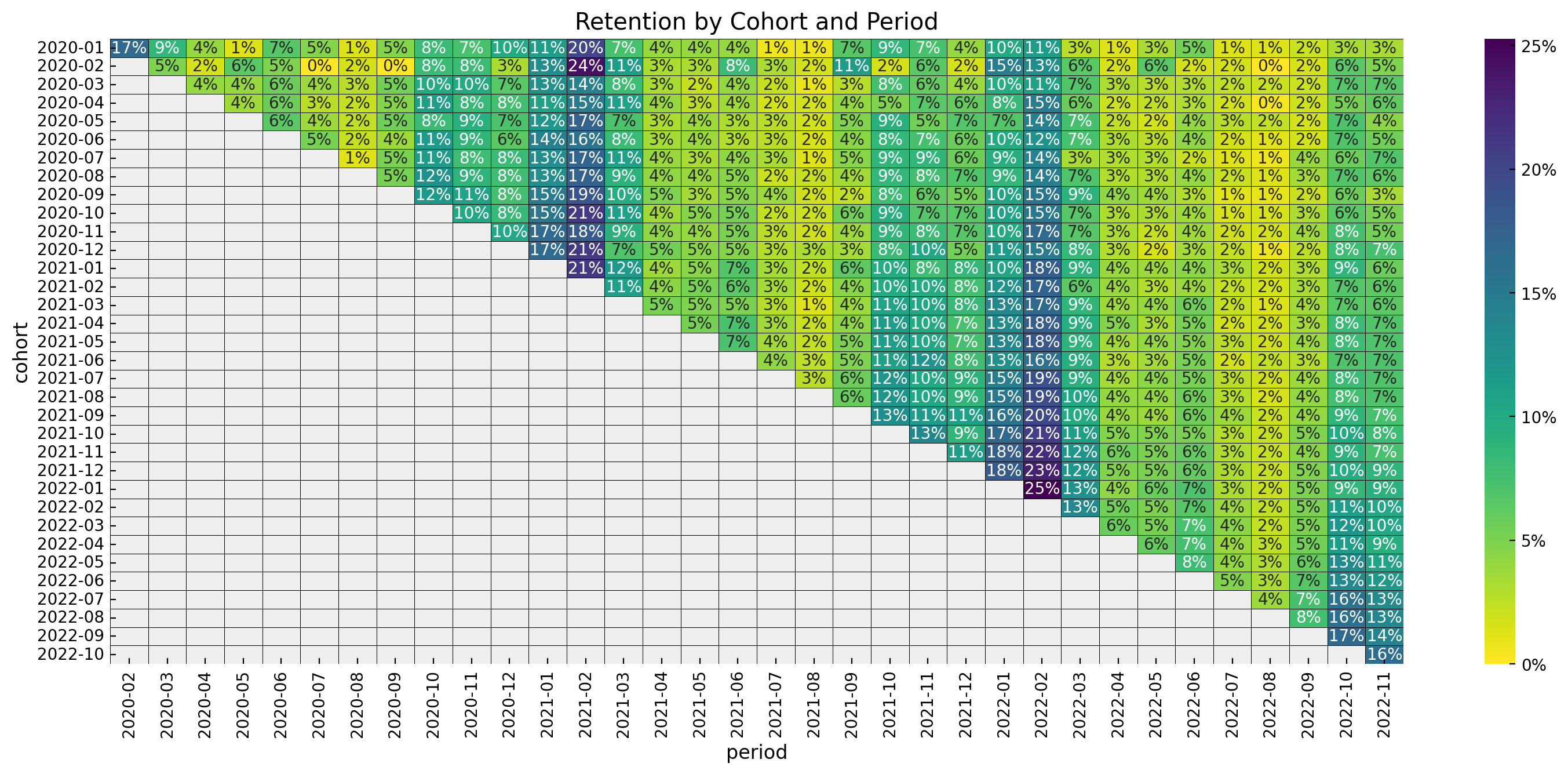}
    \caption{Retention matrix example. The matrix visualizes customer retention rates across different cohorts (rows) and
        observation periods (columns). Each cell represents the proportion of customers from a specific acquisition
        cohort that remained active in a subsequent period. Colors indicate retention rates, with darker colors typically
        showing higher retention. This visualization allows for identifying cohort-specific patterns, seasonal effects,
        and retention decay over time. The diagonal is excluded as it always contains trivial values of 1 (100\%
        retention) for the cohort's first period.}
    \label{fig:retention_matrix}
\end{figure}

This work introduces a Bayesian approach that addresses these challenges by modeling cohort-level retention rates from a
top-down perspective. Instead of building up from individual purchase patterns—a process that can become computationally
intensive and complex for large customer bases—we directly model aggregate retention and revenue at the cohort level.
to get a visual intuition of the data we want to model, Figure \ref{fig:retention_matrix} shows an example of a retention
matrix. Here we encode the cohort retention as a function of time. Note that we exclude the diagonal as it is
uninformative (always containing ones). Observe that older cohorts have more data (obviously), so we would like to use
this information to improve the estimation of retention for younger cohorts. Hence, we do not want to model each cohort
independently but rather the {\em whole retention matrix} (we will do the same for the revenue matrix and couple them
together). \\

\begin{figure}
    \centering
    \includegraphics[width=\textwidth]{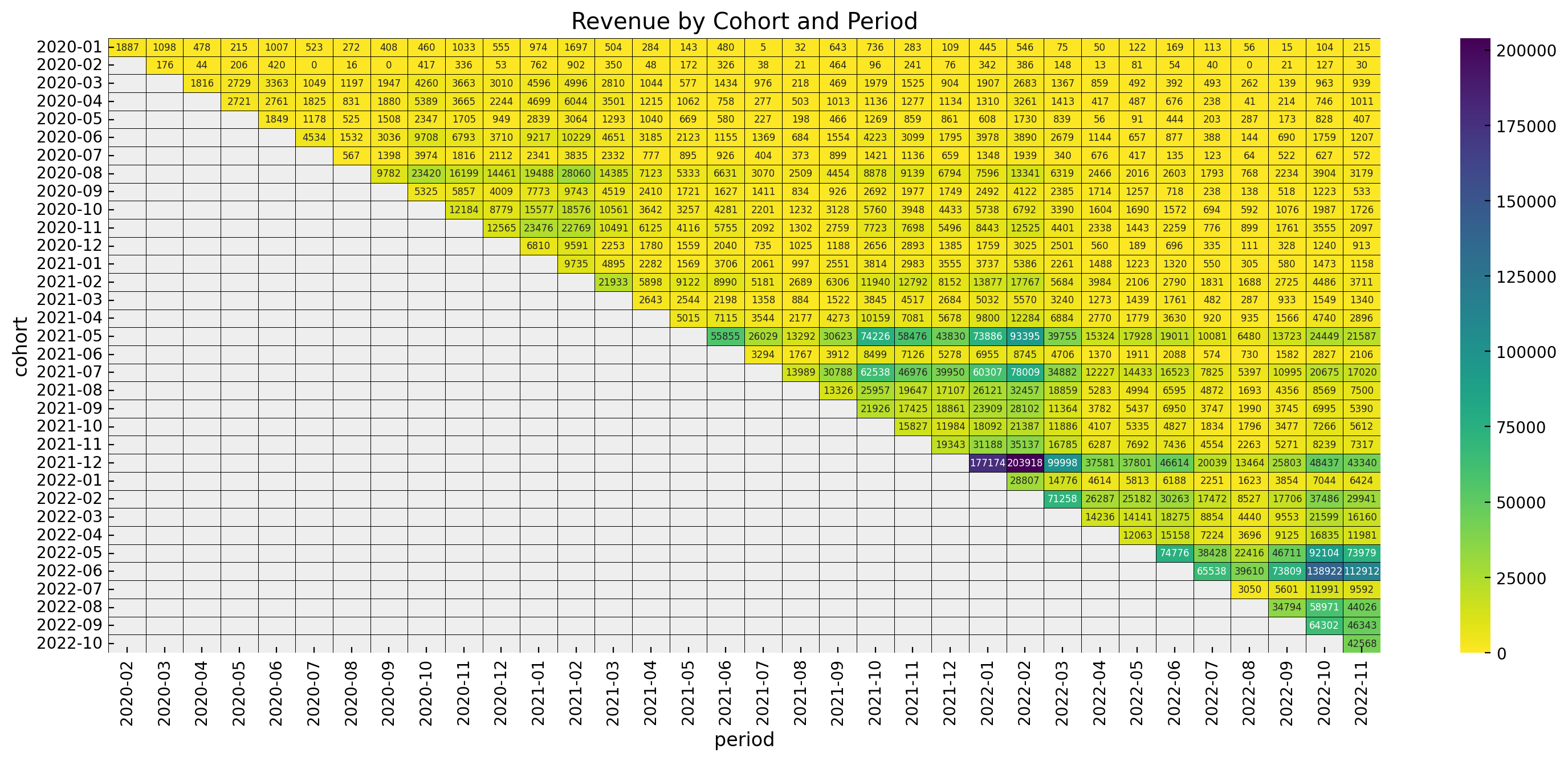}
    \caption{Revenue per cohort. This heatmap visualizes the total revenue generated by each cohort (rows) across different
        time periods (columns). The color intensity corresponds to revenue magnitude, revealing a strong correlation
        with the number of active users (Figure \ref{fig:active_users}).}
    \label{fig:revenue_matrix}
\end{figure}

In addition, as we want to understand the monetary contribution of each cohort, we can consider the revenue matrix as shown
in Figure \ref{fig:revenue_matrix}. As in the retention case, we want to make sure we use all the information available to
improve the estimation of revenue for younger cohorts. Moreover, as we will discuss below, we will couple the retention and
revenue matrices through the number of active users, making the model structure very transparent for the business users
and stakeholders.

This approach offers several distinct advantages:

\begin{itemize}
    \item \textbf{Flexibility in relationship modeling}: By employing Bayesian additive regression trees (BART)
          \cite{quiroga2022bart}, our approach can capture complex non-linear relationships between cohorts, time periods,
          and behavioral metrics without requiring explicit specification of these relationships.

    \item \textbf{Integrated seasonality}: The model naturally incorporates seasonal patterns without requiring separate
          components or preprocessing steps.

    \item \textbf{Extensibility}: Additional covariates—from macroeconomic indicators to marketing campaign intensities—can
          be seamlessly integrated into the model.

    \item \textbf{Uncertainty quantification}: The Bayesian framework provides natural uncertainty estimates around all
          predictions, enabling risk-aware decision making.

    \item \textbf{Information sharing across cohorts}: Newer cohorts with limited historical data benefit from patterns
          learned from more established cohorts.
\end{itemize}

Specifically, we use Bayesian additive regression trees to model the retention component, capturing the probability that a
customer from a given cohort remains active in subsequent periods. We couple this with a linear model for the revenue
component, predicting how much revenue active customers will generate. This dual approach balances the flexibility needed to
capture complex retention patterns with the interpretability desired for revenue forecasting. Next we describe the main
ingredients of our model: the features and the model specification. We will delve into the details in the next sections.

\subsection*{Features}

Typical purchase databases contain transactional history at user level. We want an approach general enough to benefit
from the most common features instead of heavy feature engineering. Going back to Figure \ref{fig:retention_matrix}, it
is natural to consider the following features to model the retention and revenue matrices:

\begin{itemize}
    \item {\bf Cohort age}: Age of the cohort in months, representing the time since the cohort was formed.
    \item {\bf Age}: Age of the cohort with respect to the observation time.
          This feature serves as a numerical encoder for the cohort's position in time.
    \item {\bf Month}: Month of the observation time (period), capturing seasonality effects.
\end{itemize}

For example, if our observation month is {\em 2022-11} and we consider the cohort {\em 2022-09}, the age of this cohort
is $2$ months, as the age is always calculated relative to the observation period. This cohort was observed during two
periods: {\em 2022-10} and {\em 2022-11} with cohort ages $1$ and $2$ respectively. \\

All these features are available for out-of-sample predictions, ensuring model applicability for forecasting.
In practice, we can add additional covariates to the model. The only requirement for out-of-sample predictions is that
these covariates must be available for future observation periods.

\subsection*{Model Specification}

\begin{figure}
    \centering
    \includegraphics[width=\textwidth]{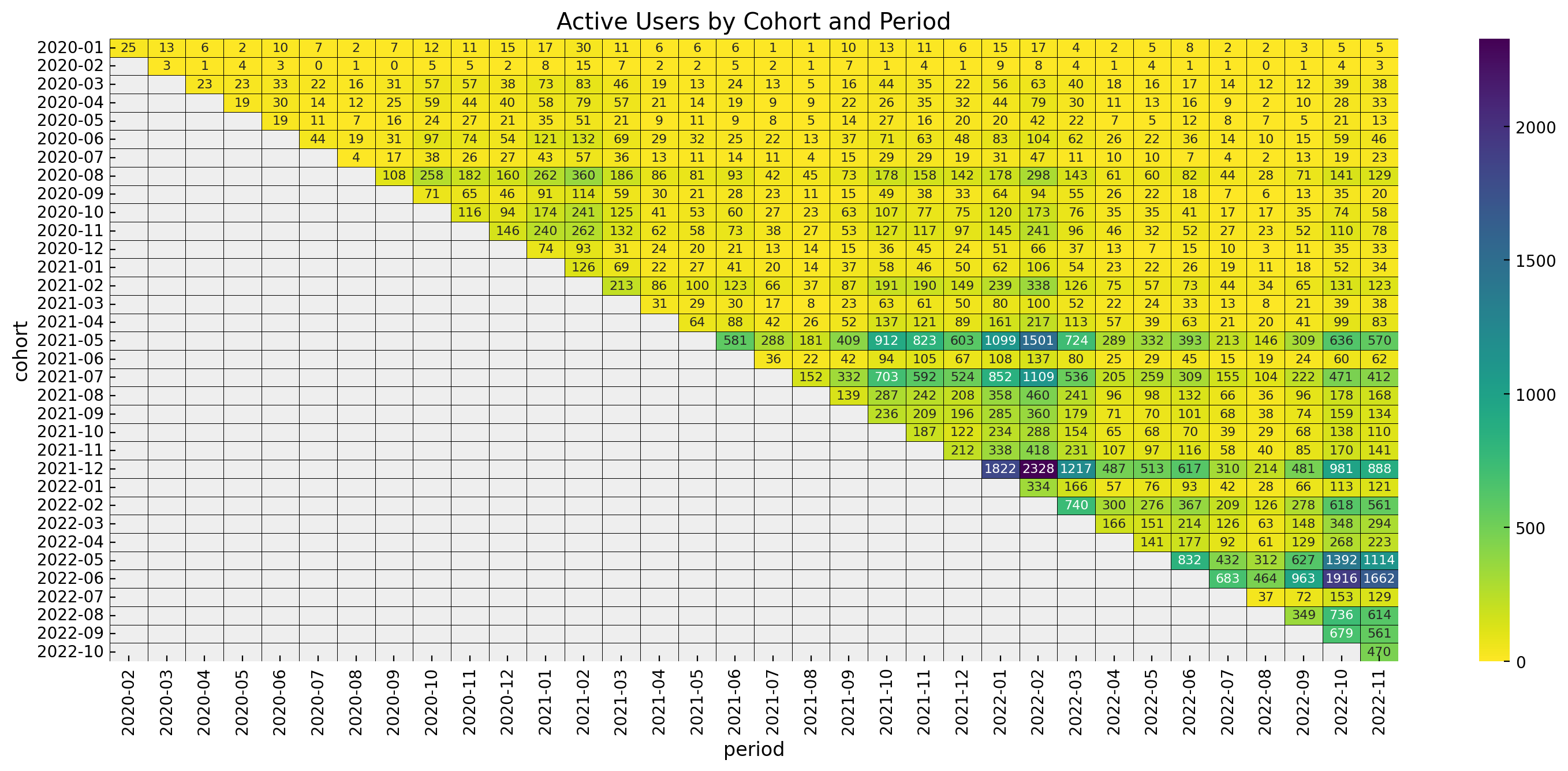}
    \caption{Number of active users across cohorts. This heatmap displays the absolute count of active users for each
        cohort (rows) across observation periods (columns).}
    \label{fig:active_users}
\end{figure}

The main idea behind the specification is to model each revenue and retention matrices, using the features above, and
couple them together. Specifically, we have:

\begin{itemize}
    \item {\bf Retention Component}: We model the number of active users $N_{\text{active}}$ in each cohort as a binomial random
          variable $\text{Binomial}(N_{\text{total}}, p)$, where the parameter $p$ represents the retention probability
          (see Figure \ref{fig:active_users}, from a synthetic example described below). We model the latent variable $p$
          using a BART model with features cohort age, age, and month. This flexible approach allows the model to
          capture non-linear relationships and interactions between features.

    \item {\bf Revenue Component}: We model the revenue matrix (see Figure \ref{fig:revenue_matrix}) through a gamma random variable
          $\text{Gamma}(N_{\text{active}}, \lambda)$, as we want to ensure non-negative values. We model the rate parameter
          $\lambda$ through a linear model with features cohort age, age, and a multiplicative interaction term (using a
          $\log$ link function). We do not explicitly add a seasonality component to this part of the model, as we typically
          observe that most seasonality effects are already captured by the retention component. However, seasonal features
          could be added if needed (plus additional features and different parametrizations, for example multiplicative
          effects).

    \item {\bf Coupling}: The retention and revenue coupling is the most interesting (and novel) part of this work. We
          couple the two components through the number of active users. Here is the full model specification:

          \begin{align*}
              \text{Revenue}    & \sim \text{Gamma}(N_{\text{active}}, \lambda)                                             \\
              \log(\lambda) = ( & \text{intercept}                                                                          \\
                                & + \beta_{\text{cohort age}} \times \text{cohort age}                                      \\
                                & + \beta_{\text{age}} \times \text{age}                                                    \\
                                & + \beta_{\text{cohort age} \times \text{age}} \times \text{cohort age} \times \text{age}) \\
              N_{\text{active}} & \sim \text{Binomial}(N_{\text{total}}, p)                                                 \\
              \textrm{logit}(p) & = \text{BART}(\text{cohort age}, \text{age}, \text{month})
          \end{align*}

\end{itemize}

\begin{figure}
    \centering
    \includegraphics[width=\textwidth]{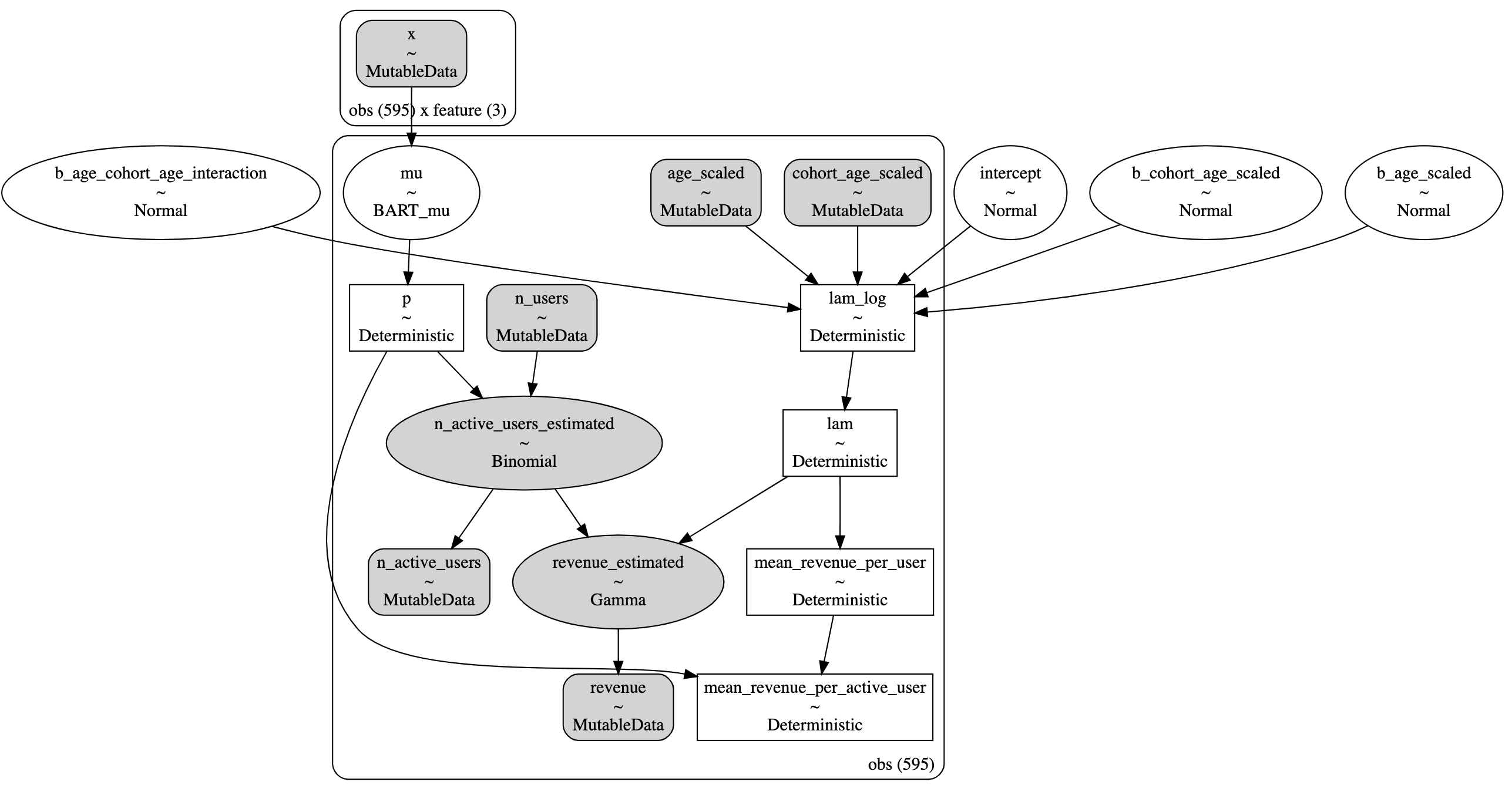}
    \caption{Cohort-revenue-retention model structure. This diagram illustrates the two coupled components of our model:
        the retention component (left) using BART to model the probability of customers remaining active, and the
        revenue component (right) using a gamma distribution with parameters informed by the retention model. The arrows
        show the flow of information, demonstrating how the estimated number of active users from the retention model
        directly feeds into the revenue model.}
    \label{fig:revenue_retention_model}
\end{figure}

Figure \ref{fig:revenue_retention_model} illustrates the complete model structure. Our goal is to simultaneously estimate
the BART parameters and the beta coefficients (including the intercept) of the linear component. We want to do this to
understand the contribution of each feature to the retention and revenue over time. Additionally, to operationalize the
model, we will use the retention and revenue matrices to make out-of-sample predictions. This can be extremely important
for scenario and business planning. A typical application is to use this model to generate {\em counterfactuals} for
global interventions where we expect different cohorts to react differently. \\

In the rest of the paper, we delve into the details of the model specification and diagnostics. Moreover, we describe
how to generate out-of-sample predictions for both the retention and revenue matrices. \\

To make the approach more tangible,
we present a synthetic dataset in the next section. This should help the reader to better understand the data and the approach.

\section{Synthetic Data}

\begin{figure}
    \centering
    \includegraphics[width=\textwidth]{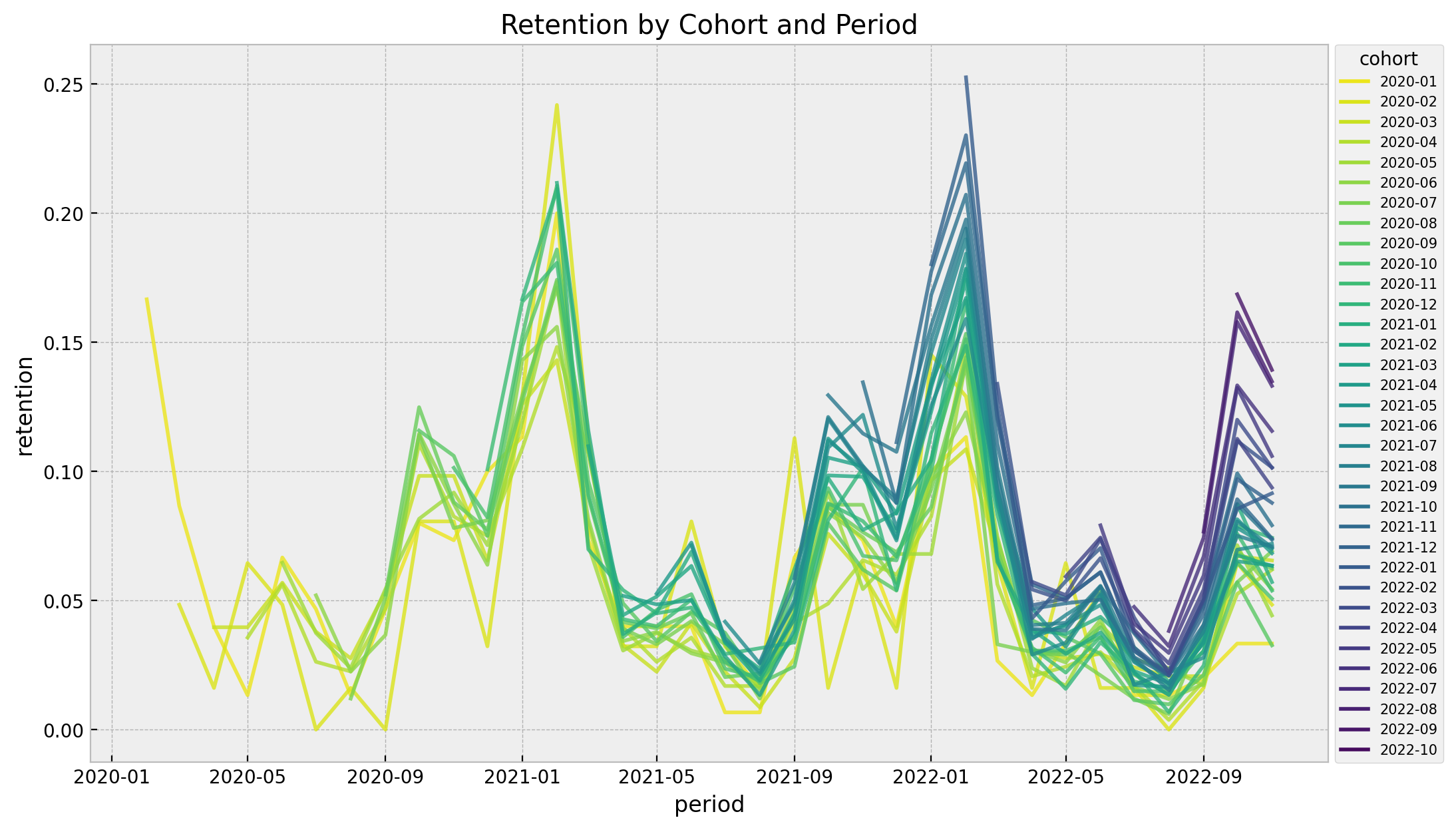}
    \caption{Retention as a function of the period, demonstrating the yearly seasonality pattern in retention values.}
    \label{fig:retention_seasonal}
\end{figure}

To illustrate our approach, we use a synthetic dataset (available as a {\em csv} file from \cite{orduz_revenue_retention}).
The code to generate this dataset (deterministically) is publicly available in \cite{orduz_revenue_retention_data_code}.
Let's begin with exploratory data analysis. Figure \ref{fig:retention_matrix} displays the retention matrix per cohort and
period. Two key observations stand out:

\begin{enumerate}
    \item The retention exhibits a clear seasonal pattern with respect to the period, being higher in the last months of the
          year and lower in the middle of the year. This seasonality pattern is more evident in Figure
          \ref{fig:retention_seasonal}.
    \item Retention appears to increase as the cohort age decreases. This trend is apparent when comparing retention values
          for periods in November across different cohort ages.
\end{enumerate}

It's important to remember that retention is a ratio, making cohort size an important factor. For instance, a retention rate
of 0.4 could represent either $4/10$ or $4\times 10^{5} / 10^{6}$. The former case carries considerably more uncertainty in
its estimation. This insight motivates us to examine the number of active users, as shown in Figure \ref{fig:active_users}.
We observe that more recent cohorts have significantly more active users, a pattern we want our model to account for. \\

\begin{figure}
    \centering
    \includegraphics[width=\textwidth]{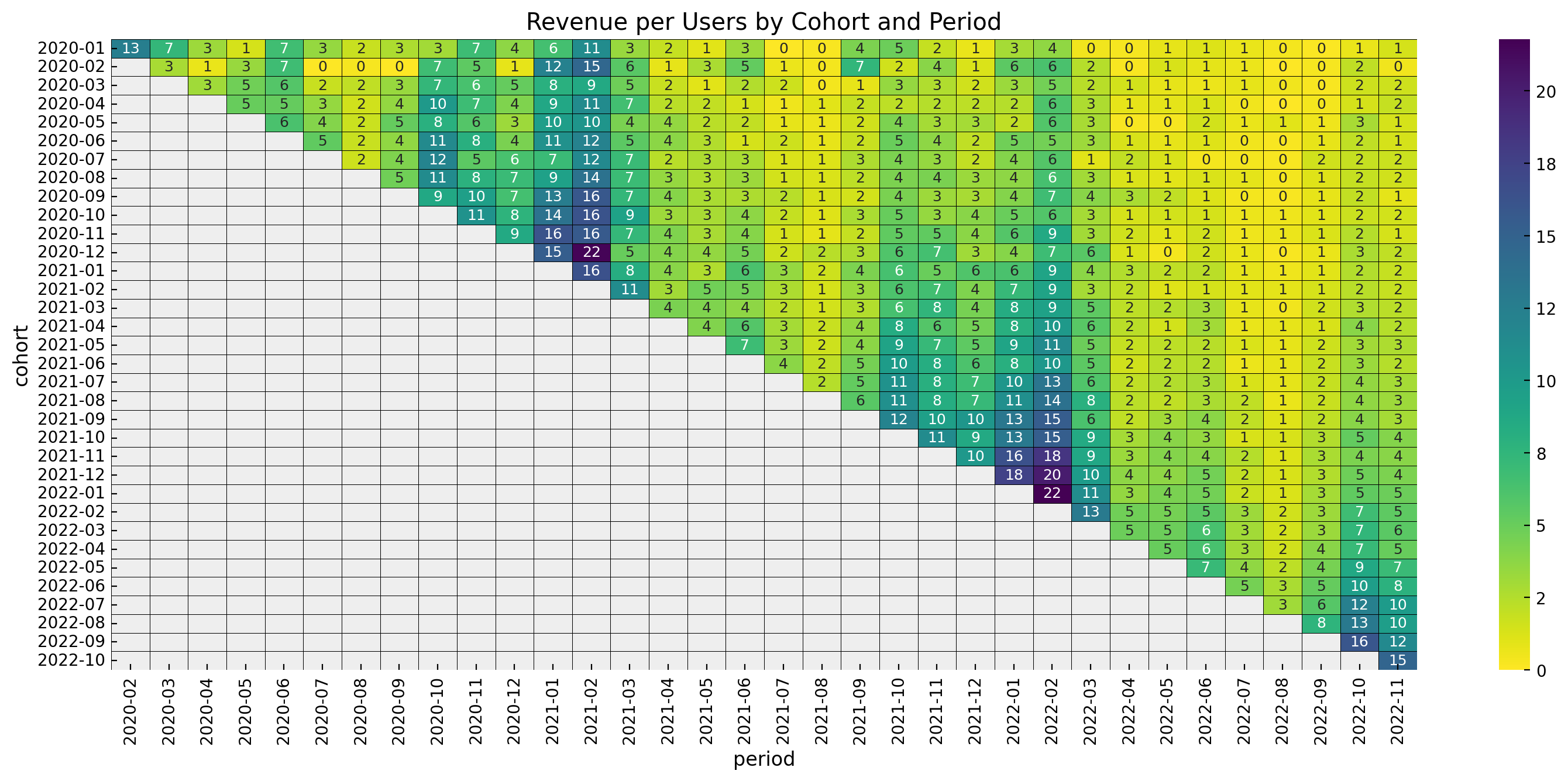}
    \caption{Revenue per user across cohorts. This visualization normalizes the total revenue by the original cohort size,
        showing the average revenue generated per initially acquired user.}
    \label{fig:revenue_per_user}
\end{figure}

\begin{figure}
    \centering
    \includegraphics[width=\textwidth]{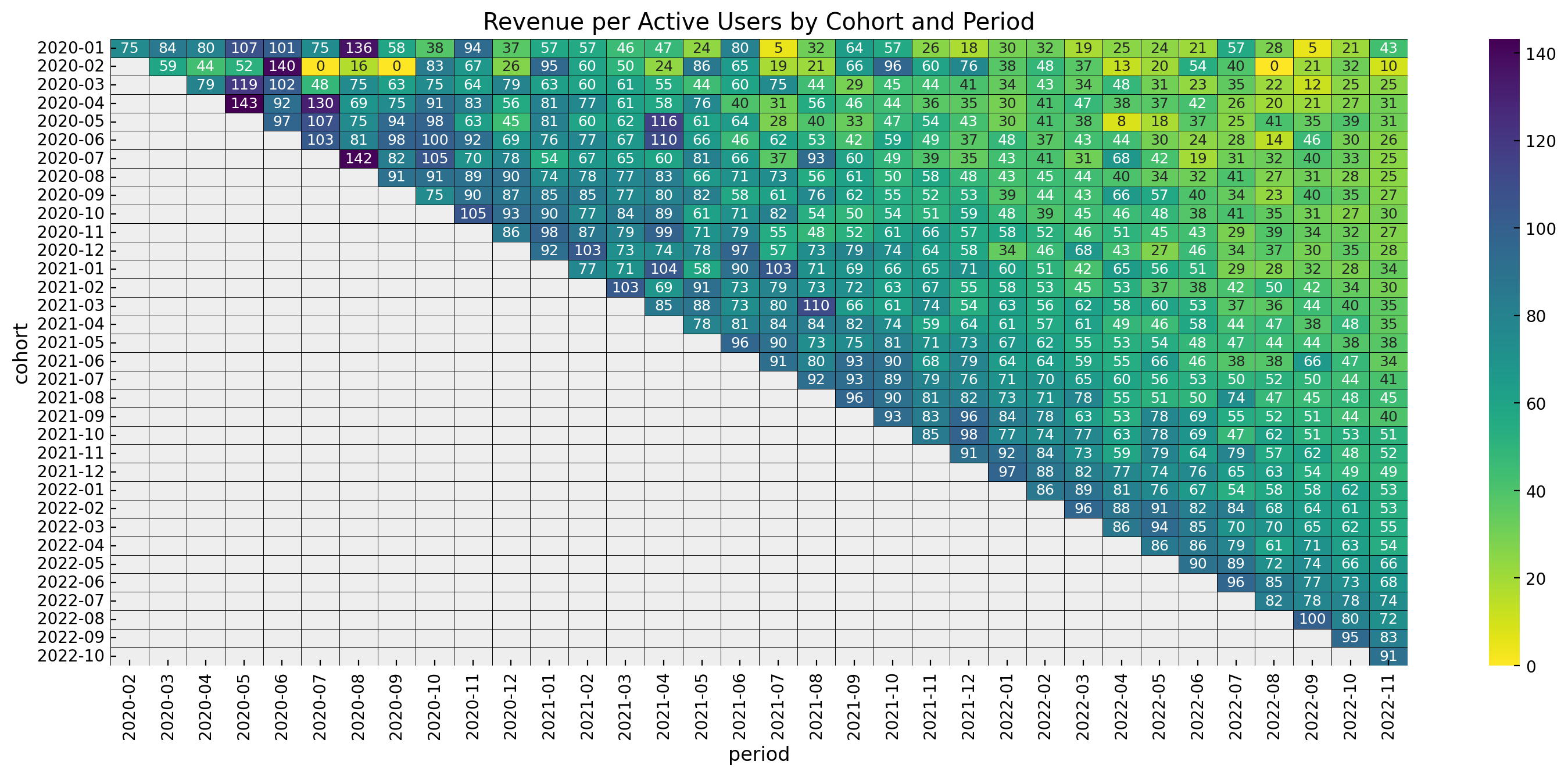}
    \caption{Revenue per active user across cohorts. This metric divides total revenue by the number of active users in each
        period, isolating spending patterns from retention effects.}
    \label{fig:revenue_per_active_user}
\end{figure}

Next, we examine revenue patterns. Figure \ref{fig:revenue_matrix} presents revenue by cohort, showing a strong correlation with the
number of active users. This suggests that revenue per user remains relatively stable over time. To verify this, we compute
revenue per user as a function of age and period (Figure \ref{fig:revenue_per_user}) as well as revenue per {\em active} user
(Figure \ref{fig:revenue_per_active_user}). The key difference between these metrics is that revenue per user divides by
total cohort size, while revenue per active user divides by the number of active users in the given period. All in all,
we observe the following for the revenue data\footnote{These types of patterns are actually common in real applications. This synthetic
    dataset is motivated by real applications where the model was proven to be very effective}:

\begin{itemize}
    \item Revenue per user exhibits a clear seasonality pattern, consistent with the seasonal pattern observed in retention.
    \item Revenue per active user does not show the same seasonality pattern since seasonal effects are already captured in
          the denominator (active users). Additionally, revenue per active user appears to decrease as cohort age increases,
          suggesting that older cohorts generate less revenue per active customer. \\
\end{itemize}

With this exploratory analysis complete, we can proceed to the modeling phase.

\section{Model Structure and Diagnostics}

\subsection{Model Specification}

Let's expand on the model structure outlined in the introduction. The core concept is to model the number of active users
as a binomial random variable $N_{\text{active}} \sim \text{Binomial}(N_{\text{total}}, p)$, where $p$ represents the retention
probability. We use Bayesian additive regression trees (BART) to model this latent variable $p$ using cohort age, age, and
month (period) as features.

\begin{align*}
    N_{\text{active}} & \sim \text{Binomial}(N_{\text{total}}, p)                  \\
    \textrm{logit}(p) & = \text{BART}(\text{cohort age}, \text{age}, \text{month})
\end{align*}

The main parameter we need to specify for the BART model is the number of trees. We typically start with a small number of
trees and increase it incrementally while monitoring the posterior predictive distribution's quality.

\begin{remark}
    A key advantage of the BART model is its flexibility in incorporating additional covariates. In real business
    applications, we have successfully added customer segmentation features (such as acquisition media channels from
    attribution models). This provides valuable insights into media channel return-on-investment (ROI), allowing businesses
    to consider not just acquisition costs but also estimated customer lifetime value through this combined model.
\end{remark}

\begin{remark}
    While one could start with a simpler model, such as a linear model as described in \cite{orduz_revenue_retention},
    our experience with real datasets shows that such simpler approaches often fail to adequately capture the complex
    patterns in the data.
\end{remark}

For the revenue component, we employ a gamma random variable $\text{Gamma}(N_{\text{active}}, \lambda)$ (inspired by
\cite{stucchio2015bayesian}). The mean of this gamma distribution is $N_{\text{active}} / \lambda$, allowing us to interpret
$1 / \lambda$ as the {\em average revenue per active user}. We model $\log(\lambda)$ using a linear function of cohort age,
age, and their interaction.

\begin{align*}
    \text{Revenue}    & \sim \text{Gamma}(N_{\text{active}}, \lambda)                                             \\
    \log(\lambda) = ( & \text{intercept}                                                                          \\
                      & + \beta_{\text{cohort age}} \times \text{cohort age}                                      \\
                      & + \beta_{\text{age}} \times \text{age}                                                    \\
                      & + \beta_{\text{cohort age} \times \text{age}} \times \text{cohort age} \times \text{age})
\end{align*}

A key insight from both this synthetic dataset and many real-world applications is that we typically don't need to explicitly
model seasonality in the revenue component, as seasonal patterns are already captured by the retention component.

\begin{remark}
    The {\em age} feature characterizes each cohort's temporal position. While we could replace this numerical encoding with
    a one-hot encoding of cohorts and add hierarchical structure to pool information across cohorts, the numerical encoding
    is more parsimonious under the assumption that temporally proximate cohorts behave more similarly than distant ones.
\end{remark}

As a preprocessing step, we standardize the features for the linear model component (we keep the same notation for the
variables for simplicity). This allows us to specify priors for the regression coefficients in terms of the effect of a
one-standard-deviation change in the predictor, enabling effective regularization through standard normal priors for the
coefficients (see \cite{orduz_retention_bart}). \\

In summary, our cohort-revenue-retention model is specified as:

\begin{align*}
    \text{Revenue}                              & \sim \text{Gamma}(N_{\text{active}}, \lambda)                                             \\
    \log(\lambda) = (                           & \text{intercept}                                                                          \\
                                                & + \beta_{\text{cohort age}} \times \text{cohort age}                                      \\
                                                & + \beta_{\text{age}} \times \text{age}                                                    \\
                                                & + \beta_{\text{cohort age} \times \text{age}} \times \text{cohort age} \times \text{age}) \\
    N_{\text{active}}                           & \sim \text{Binomial}(N_{\text{total}}, p)                                                 \\
    \textrm{logit}(p)                           & = \text{BART}(\text{cohort age}, \text{age}, \text{month})                                \\
    \text{intercept}                            & \sim \text{Normal}(0, 1)                                                                  \\
    \beta_{\text{cohort age}}                   & \sim \text{Normal}(0, 1)                                                                  \\
    \beta_{\text{age}}                          & \sim \text{Normal}(0, 1)                                                                  \\
    \beta_{\text{cohort age} \times \text{age}} & \sim \text{Normal}(0, 1)
\end{align*}

\subsection{Diagnostics}

\begin{figure}
    \centering
    \includegraphics[width=\textwidth]{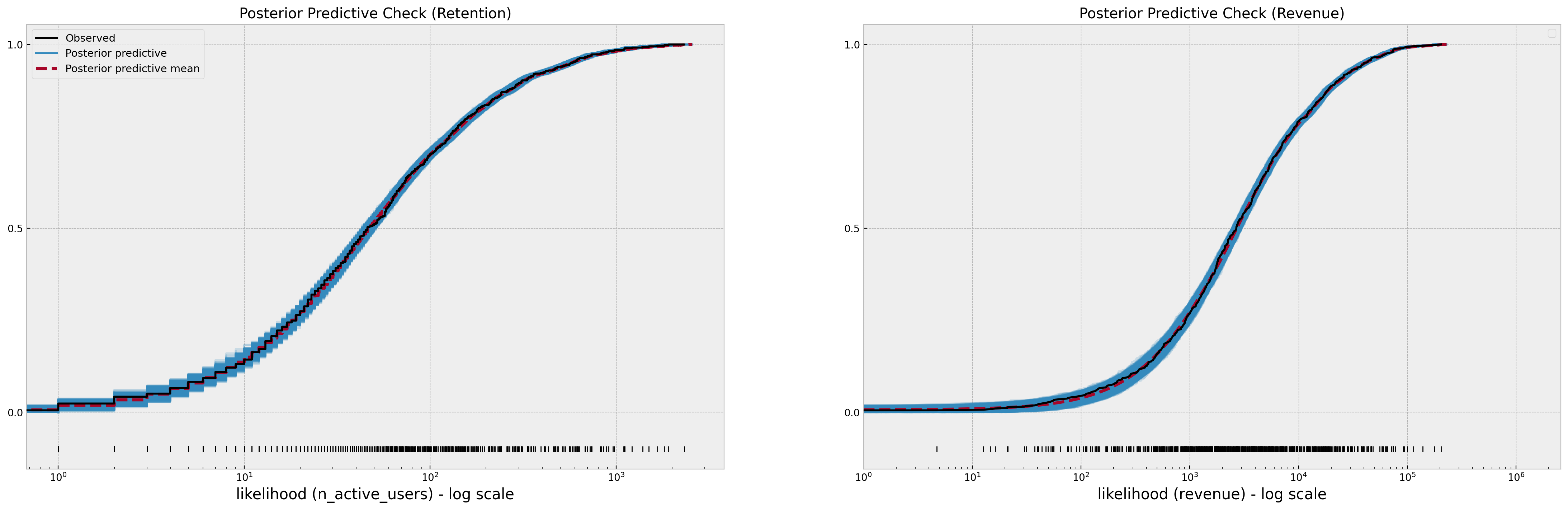}
    \caption{Posterior predictive distribution of the retention (left) and revenue (right) components, showing good fit to
        the observed data. These cumulative density plots compare the distributions of observed values (black) with simulated
        values from the posterior predictive distribution (orange), providing a visual assessment of model fit.}
    \label{fig:posterior_predictive}
\end{figure}

\begin{figure}
    \centering
    \includegraphics[width=\textwidth]{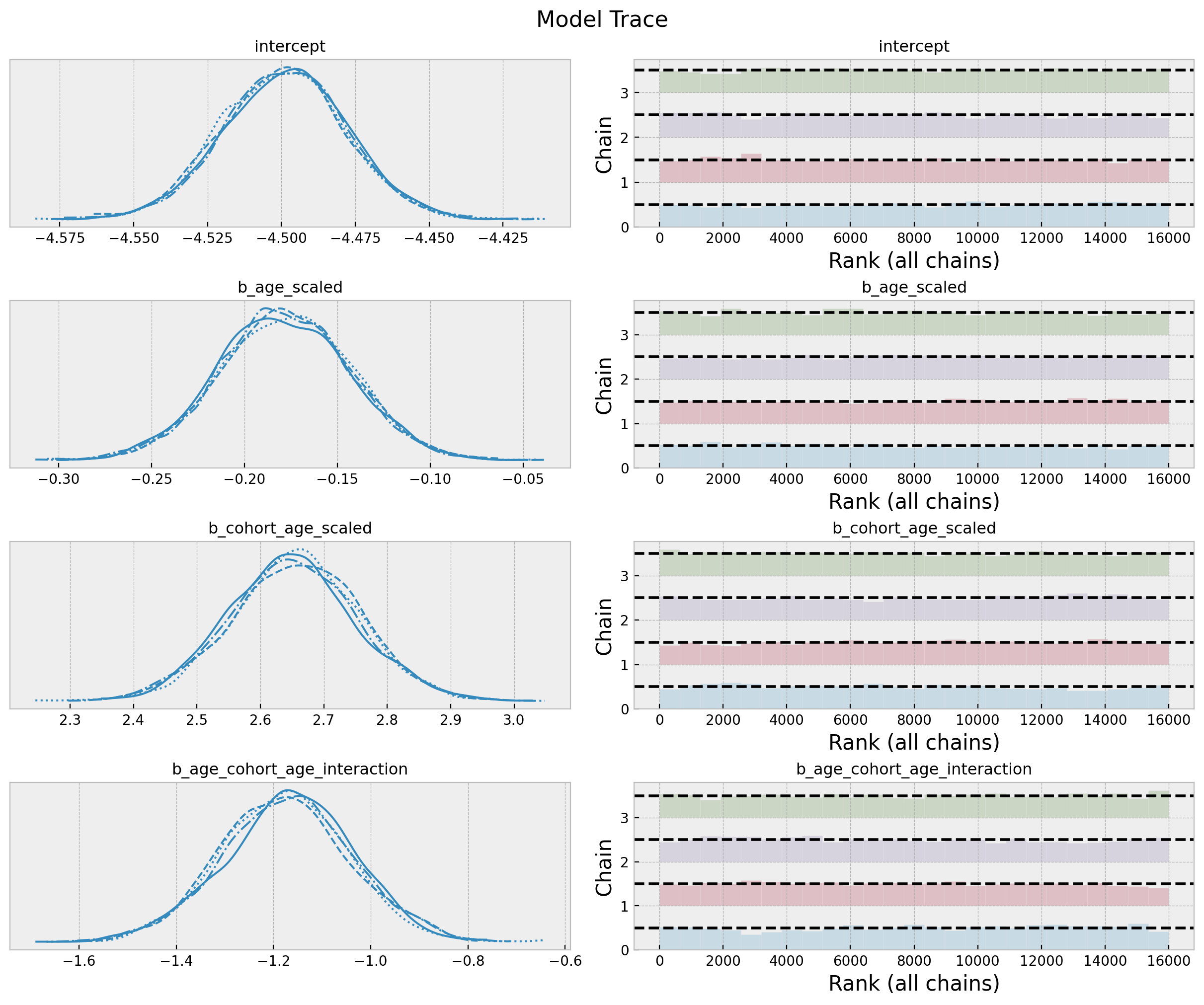}
    \caption{Trace plots for the linear model parameters, showing good mixing and convergence of the MCMC chains.}
    \label{fig:trace}
\end{figure}

Once we have the model specification, we can implement and fit it in PyMC (see \cite{orduz_revenue_retention}). Figure
\ref{fig:posterior_predictive} shows the posterior predictive distribution of both model components. The trace plots for the
linear terms (Figure \ref{fig:trace}) show good mixing with no divergences or convergence warnings.

\subsection{Variable Importance}

\begin{figure}
    \centering
    \includegraphics[width=\textwidth]{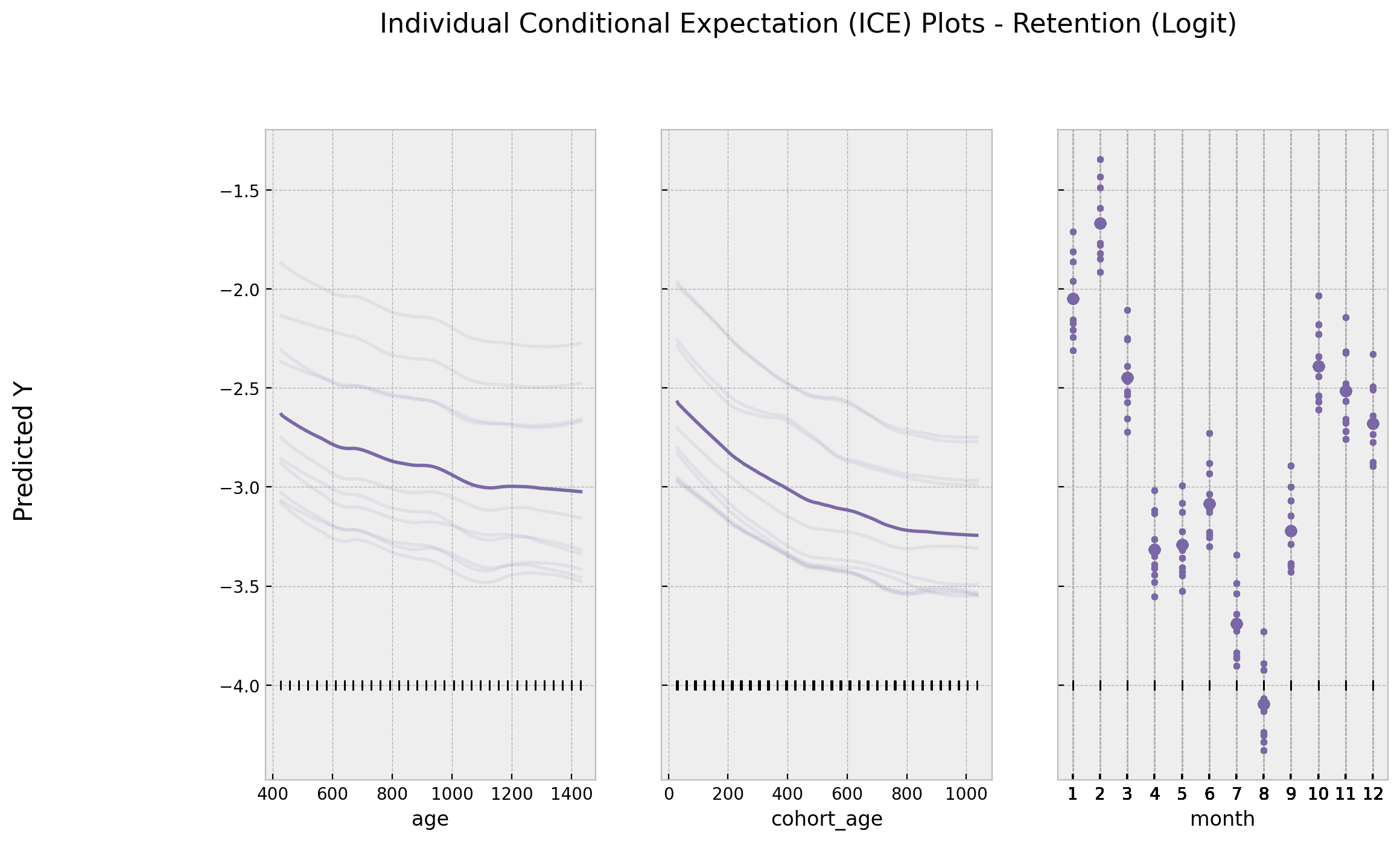}
    \caption{PDP (solid line) and ICE (dashed lines) plots for the retention component.}
    \label{fig:ice_plot}
\end{figure}

The BART component implemented in \cite{quiroga2022bart} provides great tools to understand the importance of the different
features. Figure \ref{fig:ice_plot} shows the PDP (Partial Dependence Plot) and ICE (Individual Conditional Expectation)
plots for the retention component. PDP and ICE plots visualize how the model's predictions change as a single feature varies
while holding all other features constant. Each line represents a different observation from the dataset, showing how the
predicted retention probability would change for that observation if we modified only the feature of interest. The PDP
plot is the average of the ICE plots (solid line). These plots allow us to understand how the retention probability varies
for different values of the features and reveals potential non-linear relationships or interaction effects that might not
be apparent in aggregate statistics. \\

In this specific example, we can extract the following insights:

\begin{itemize}
    \item The ICE plots show how the retention rate decreases with both cohort age and age. This is not surprising as we saw in the EDA.
    \item We see that the ICE plots have a similar trend to the PDP plots. This hints that the interaction effects are
          not so important in this case. This is also something we saw in the linear model where the interaction coefficient
          was relatively small (see \cite{orduz_retention}).
    \item We clearly see the seasonality component of the PDP / ICE plots resemble the regression coefficients in the linear
          model from \cite{orduz_retention}. This is simply representing the strong seasonal component of the data.
\end{itemize}

\begin{figure}
    \centering
    \includegraphics[width=\textwidth]{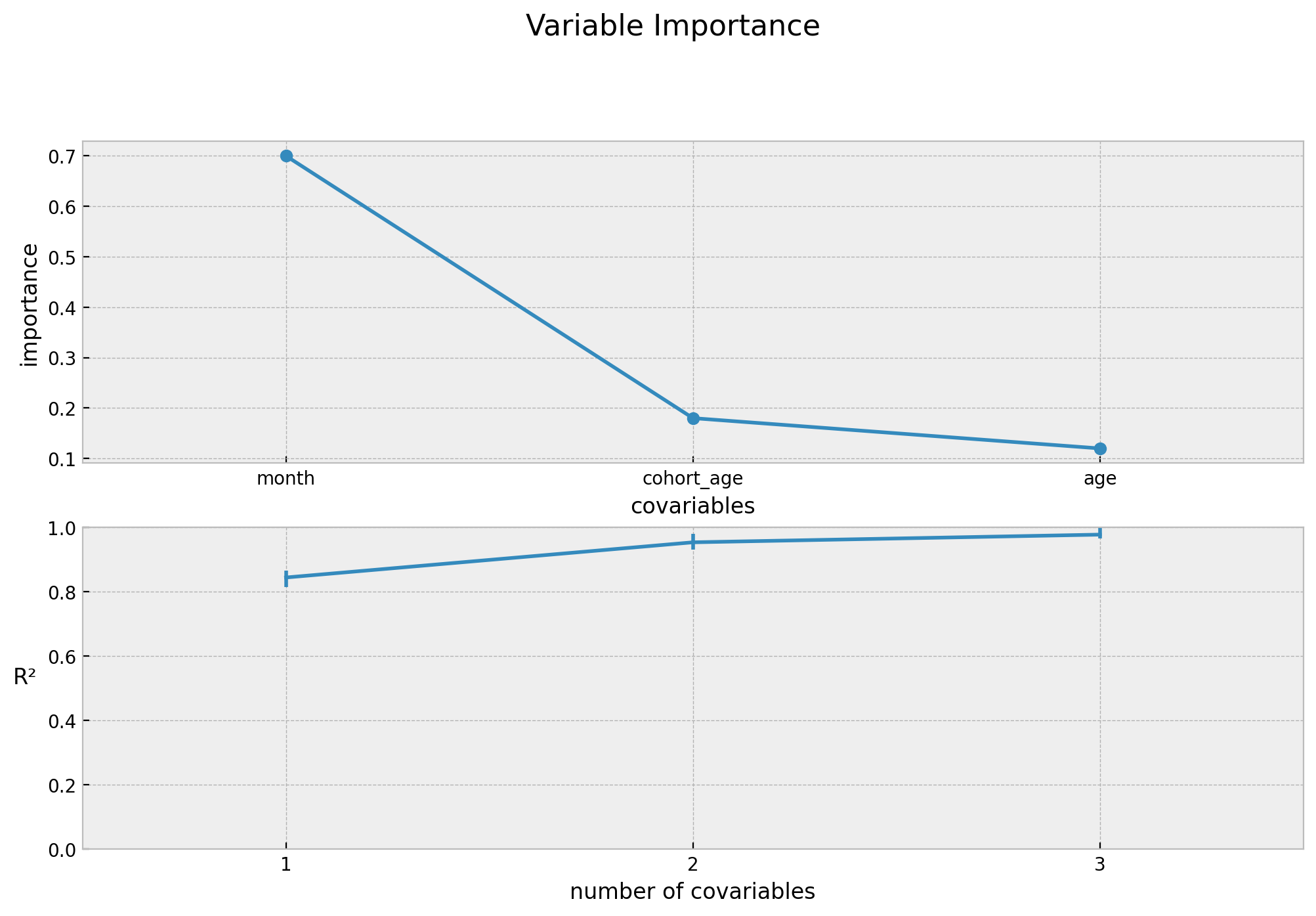}
    \caption{Variable importance for the retention component based on the in-sample $R^2$.}
    \label{fig:variable_importance}
\end{figure}

In addition, we can extract a relative importance for the different features using the contribution to the in-sample $R^2$,
as shown in Figure \ref{fig:variable_importance}.

These types of plots are very valuable to understand the {\em drivers} of the retention component.

\section{Predictions}

In this section, we present both in-sample and out-of-sample predictions from our model, demonstrating its effectiveness at
capturing patterns in the data and forecasting future metrics.

\subsection{In-Sample Predictions}

\begin{figure}
    \centering
    \begin{tabular}{cc}
        \includegraphics[width=0.5 \textwidth]{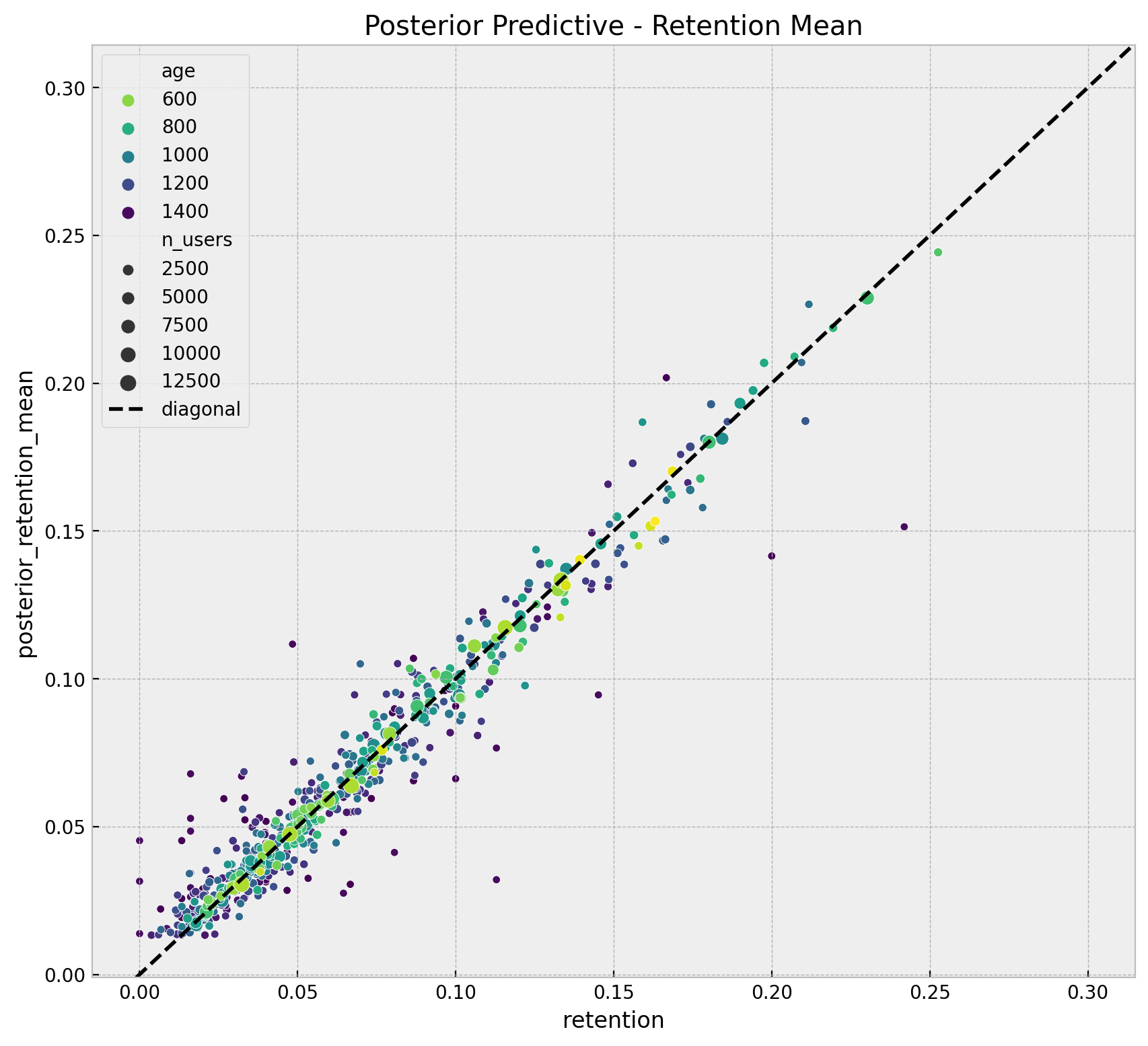} &
        \includegraphics[width=0.5 \textwidth]{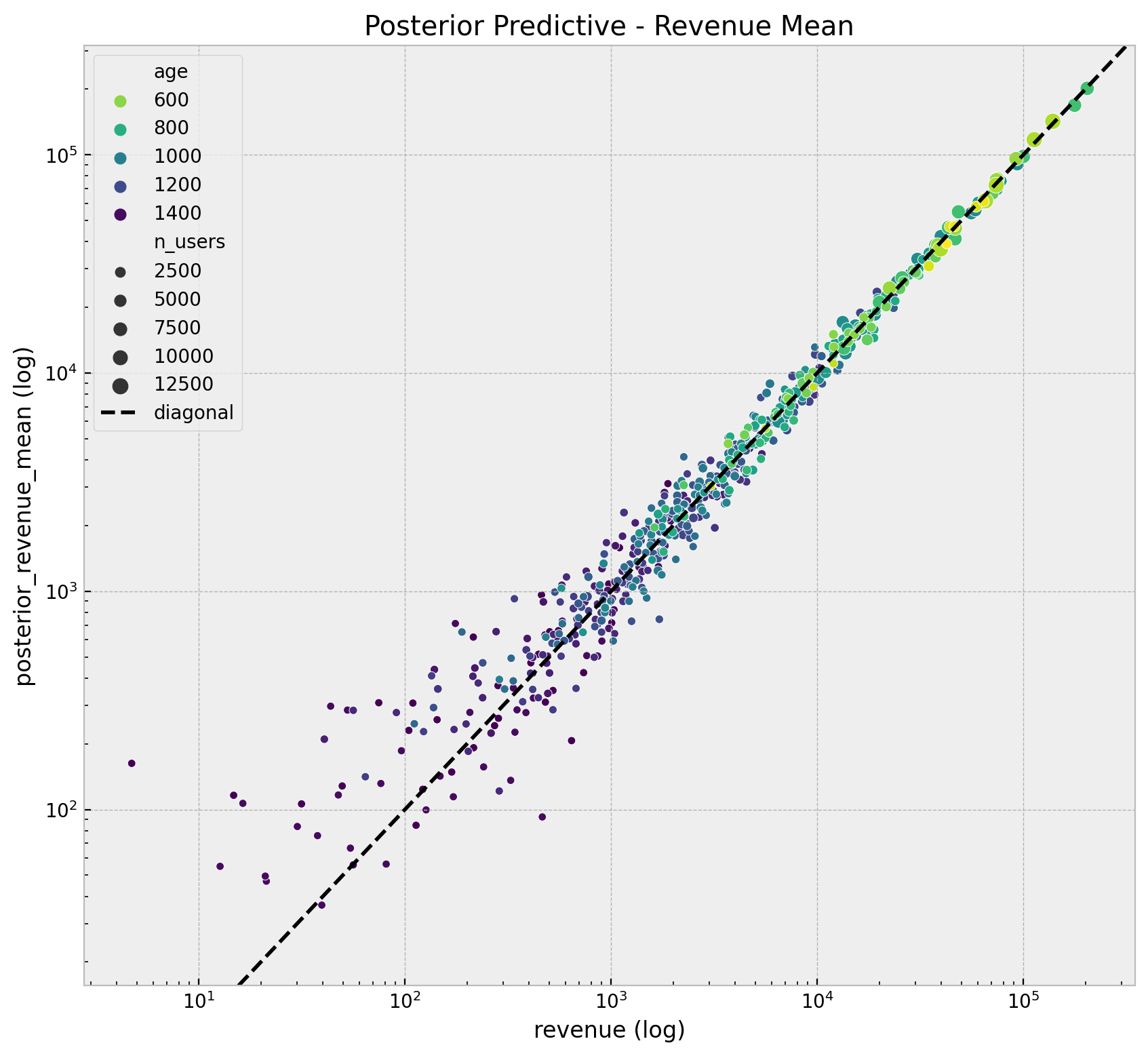}
    \end{tabular}
    \caption{Retention (left) and revenue (right) in-sample posterior predictive mean values plotted against the actual
        observations. These scatter plots provide a quantitative assessment of model fit by comparing predicted
        versus observed values, with points closer to the diagonal line indicating better predictions.}
    \label{fig:in_sample_mean}
\end{figure}

\begin{figure}
    \centering
    \includegraphics[width=\textwidth]{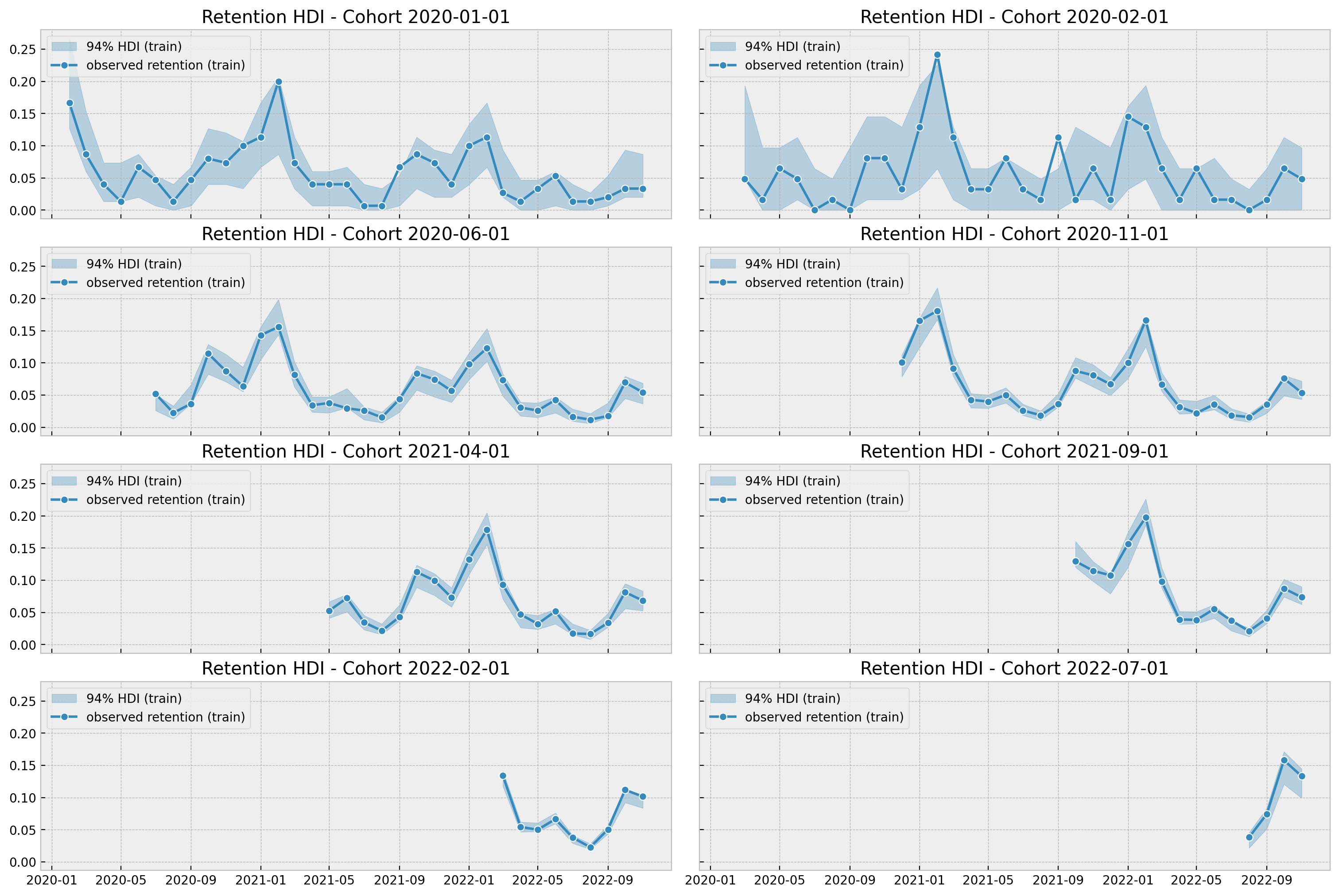}
    \caption{Retention in-sample posterior predictive distribution for selected cohorts, showing $94\%$ HDI (blue shaded areas)
        and observed retention values (blue points). This visualization displays the model's predictive performance for
        retention across time for different cohorts, with uncertainty quantified through highest density intervals.
        The narrower intervals for more recent cohorts (bottom panels) reflect greater certainty due to more available
        data, while the consistent capture of observed values within the intervals indicates well-calibrated uncertainty
        estimates. The plots also reveal the model's ability to adapt to cohort-specific patterns and seasonal
        fluctuations, demonstrating its flexibility in capturing complex temporal dynamics.}
    \label{fig:in_sample_retention}
\end{figure}

\begin{figure}
    \centering
    \includegraphics[width=\textwidth]{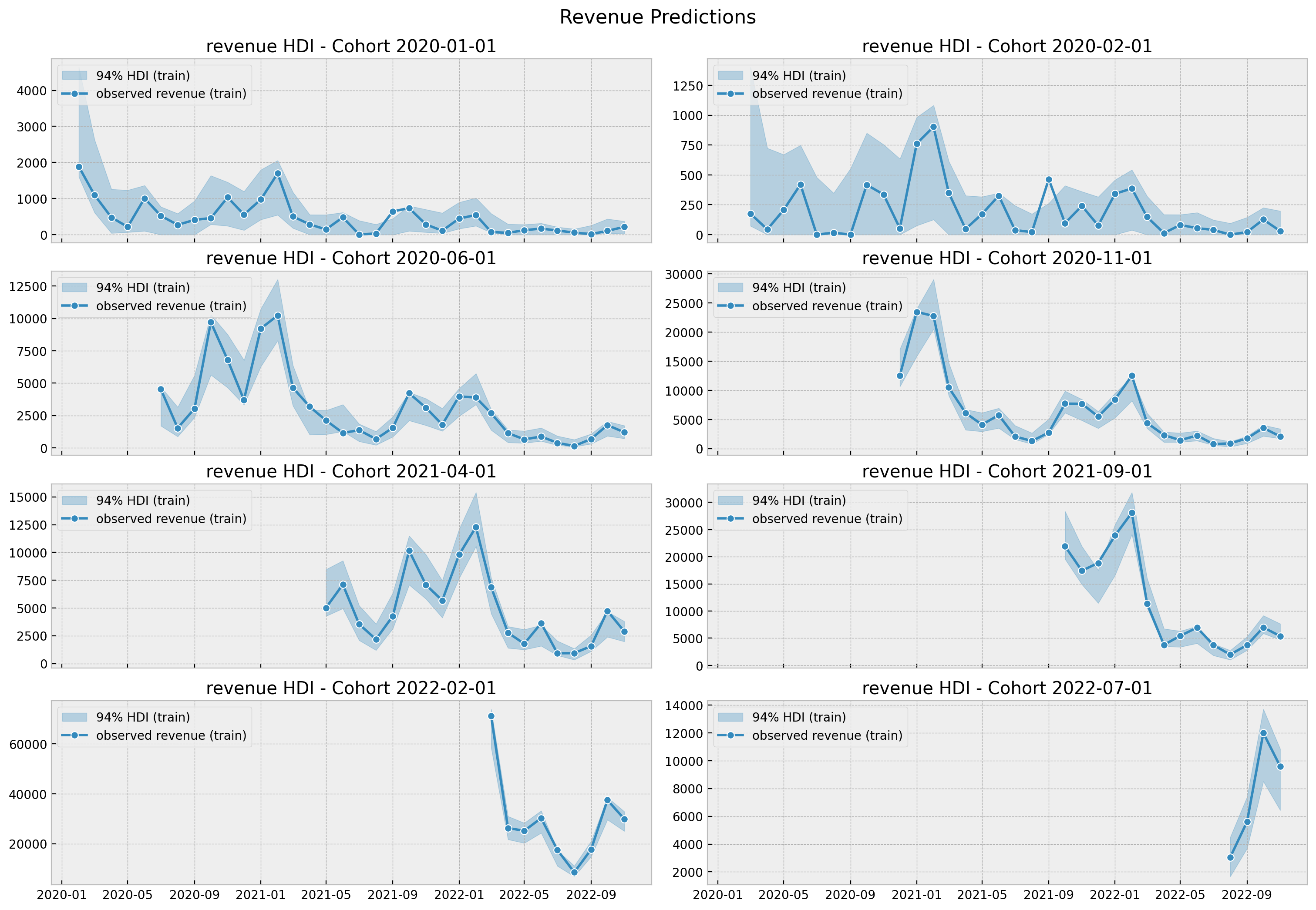}
    \caption{Revenue in-sample posterior predictive distribution for selected cohorts, showing $94\%$ HDI (blue shaded areas)
        and observed revenue values (blue points). These plots illustrate the model's revenue predictions and associated
        uncertainty across time for different cohorts. The successful capture of observed values within the HDI bands
        demonstrates the model's ability to accurately represent not just central tendencies but also the inherent
        variability in revenue. The visualization highlights how our coupled modeling approach effectively propagates
        uncertainty from the retention component to revenue estimates, providing business stakeholders with realistic
        confidence intervals for financial planning and analysis.}
    \label{fig:in_sample_revenue}
\end{figure}

\begin{figure}
    \includegraphics[width=\textwidth]{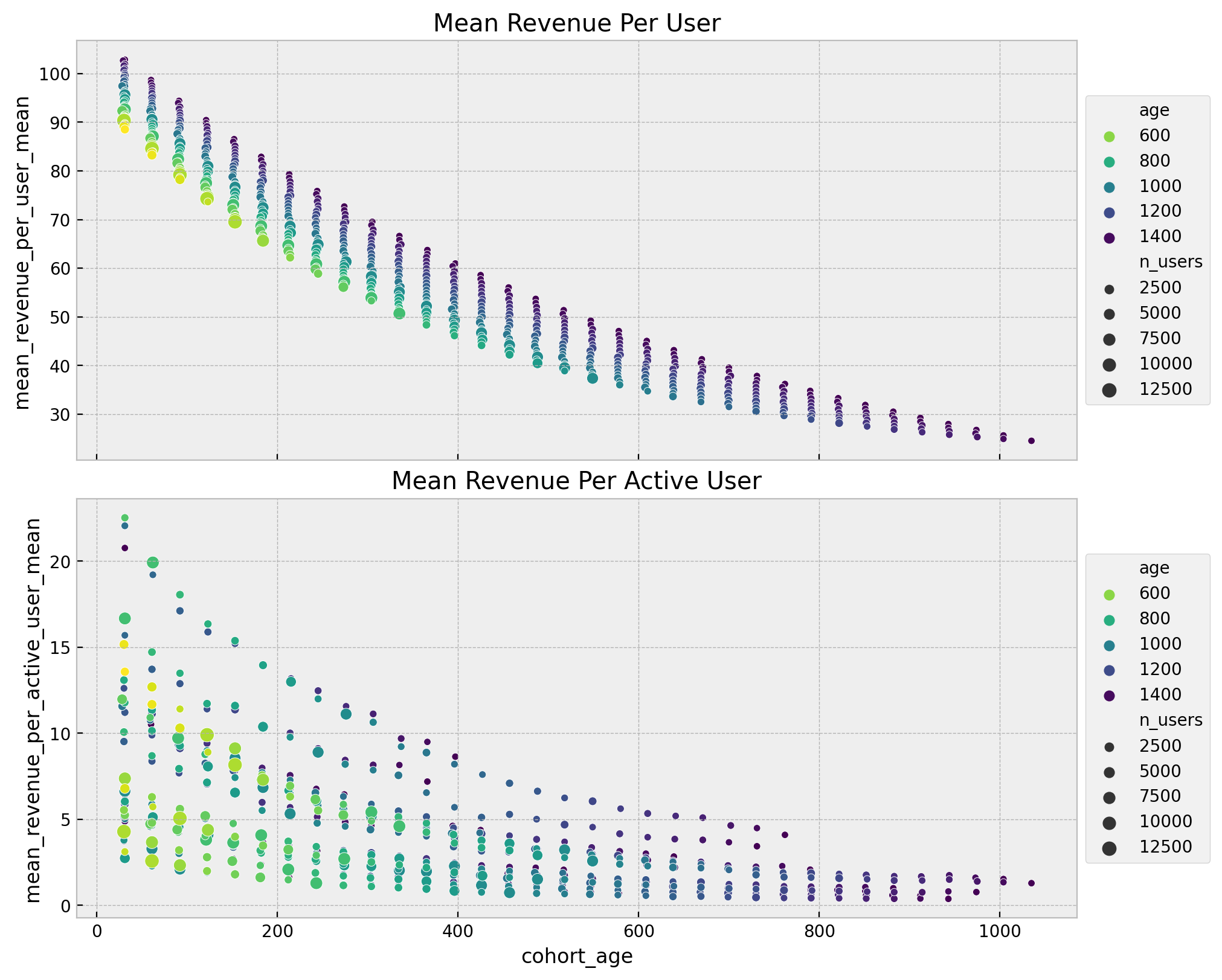}
    \caption{Additional view of posterior predictions across cohorts, illustrating the model's ability to capture
        cohort-specific patterns. This panel view organizes predictions by cohort (columns) and shows how the model
        adapts to the unique characteristics of each customer group.}
    \label{fig:additional_predictions}
\end{figure}

We first evaluate the model's in-sample performance by comparing the posterior predictive mean against the observed values.
Figure \ref{fig:in_sample_mean} shows the comparison for both retention and revenue components, with points closer to the
diagonal line indicating better fit. Beyond point estimates, we can visualize the full posterior predictive distribution
to assess model uncertainty. Figure \ref{fig:in_sample_retention} shows the posterior predictive distribution of retention
for selected cohorts, with $94\%$ HDI (Highest Density Interval). Note how the intervals are narrower for more recent
cohorts with more data, reflecting greater certainty in these predictions. Overall, the predictions effectively capture the
observed retention patterns, including seasonality. For the revenue component, Figure \ref{fig:in_sample_revenue} shows
the posterior predictive distribution compared to actual revenue values. The model successfully captures the revenue
variability across different cohorts and time periods. We can use the whole posterior distribution to make custom visualizations
of quantities of interests like the revenue per active user, as shown in Figure \ref{fig:additional_predictions}.

\subsection{Out-of-Sample Predictions}

\begin{figure}
    \centering
    \includegraphics[width=\textwidth]{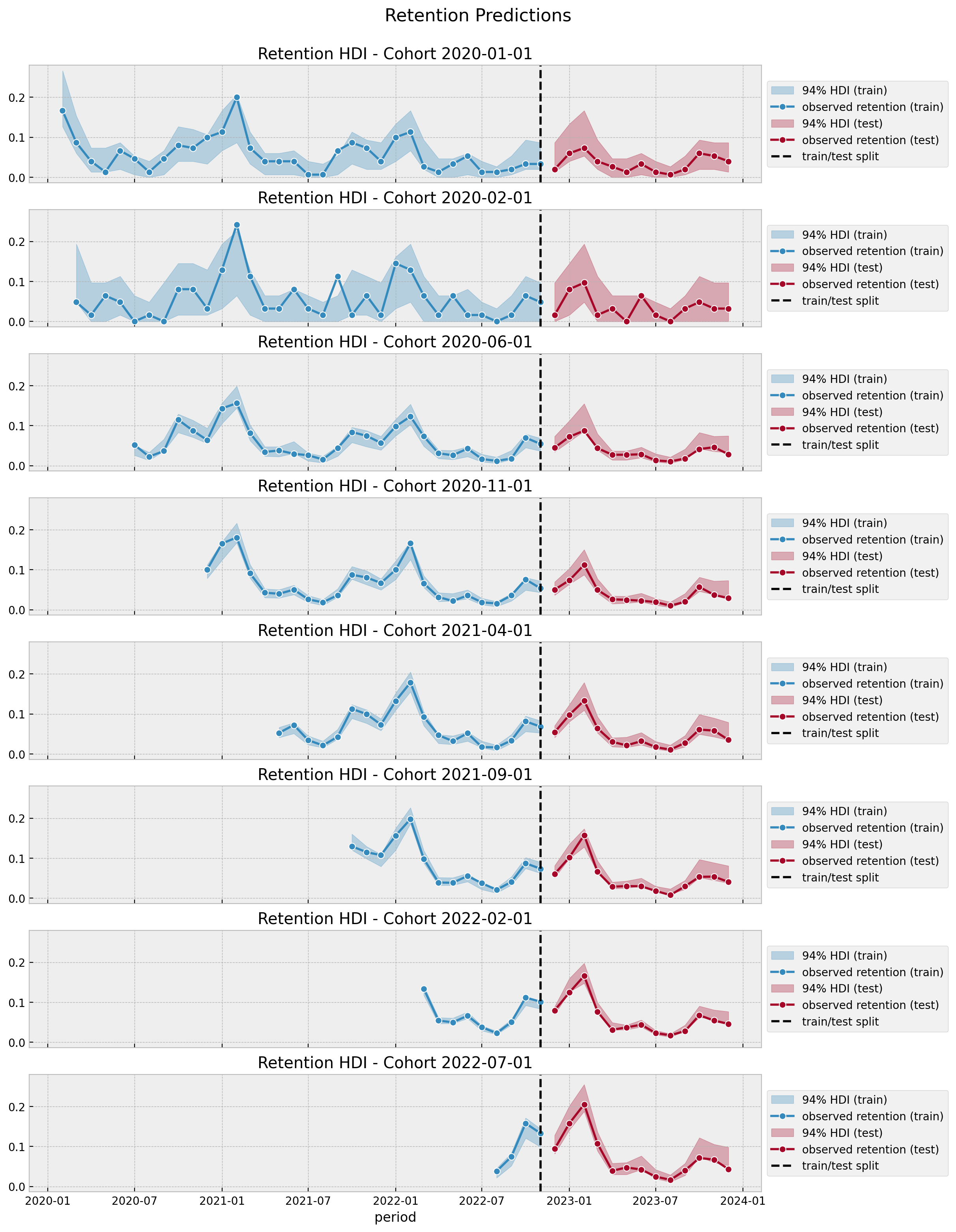}
    \caption{Retention out-of-sample posterior predictive distribution for (random) selected cohorts.}
    \label{fig:out_sample_retention}
\end{figure}

\begin{figure}
    \centering
    \includegraphics[width=\textwidth]{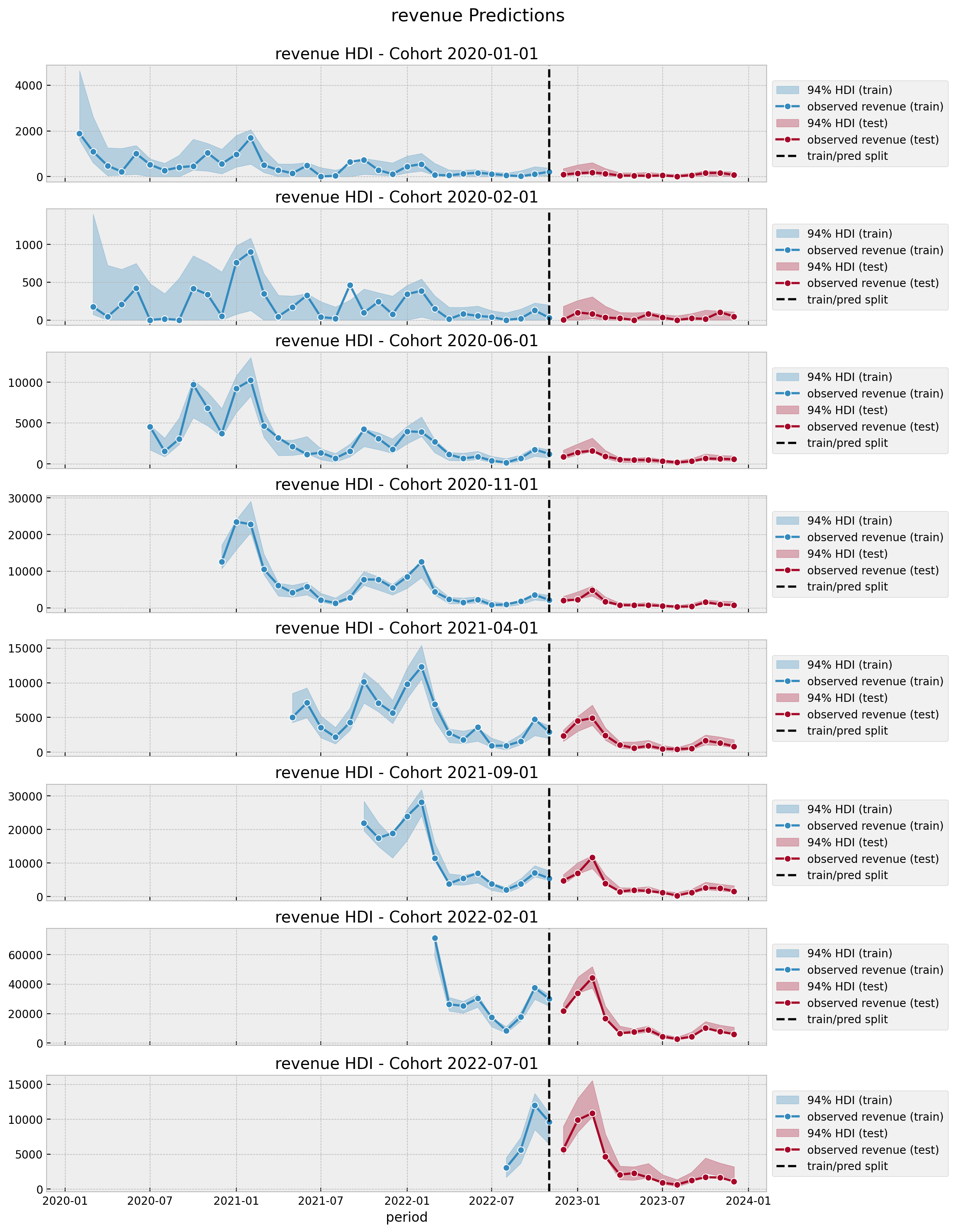}
    \caption{Revenue out-of-sample posterior predictive distribution for (random) selected cohorts.}
    \label{fig:out_sample_revenue}
\end{figure}

The true test of any predictive model is its performance on unseen data. We evaluate our model's forecasting capabilities
using a holdout set consisting of data after {\em 2022-11}, which was not used during model training. Figures
\ref{fig:out_sample_retention} and \ref{fig:out_sample_revenue} show the out-of-sample predictions for retention and
revenue, respectively. The vertical dashed lines indicate the train/test split point. Several key observations emerge:

\begin{enumerate}
    \item The model successfully predicts both retention and revenue patterns for future periods, with most actual
          observations falling within the $94\%$ HDI.

    \item The model effectively captures the seasonal patterns in retention, correctly predicting the expected peaks and
          troughs in future months based on historical patterns.

    \item For newer cohorts with limited training data (e.g., the {\em 2022-07} cohort with only 4 data points in
          training), the model still produces reasonable predictions by leveraging information learned from older cohorts.
          This demonstrates effective transfer of knowledge across cohorts.

    \item The $94\%$ HDI appropriately widens for more distant future predictions, reflecting increasing uncertainty as we
          forecast further ahead. \\
\end{enumerate}

These results highlight one of the key advantages of our Bayesian approach: the ability to make probabilistic forecasts with
well-calibrated uncertainty using highest density intervals (HDI). The model provides not just point estimates but complete
distributions, allowing businesses to understand the range of possible outcomes and make risk-aware decisions. The effective
transfer of information across cohorts is particularly valuable for new cohorts where limited data is available.

\section{Other Non-Parametric Approaches}

While Bayesian Additive Regression Trees (BART) provide a powerful non-parametric approach for modeling the retention
component, there are other flexible methods worth considering. In particular, neural networks coupled with efficient
Bayesian inference techniques offer an alternative that combines flexibility with computational efficiency.

\subsection{Neural Networks with NumPyro}

As demonstrated by \cite{orduz_revenue_retention_numpyro}, the BART component in our model can be replaced with a neural
network implemented using Flax, with inference performed using NumPyro \cite{phan2019composable}. The modified model
structure becomes:

\begin{align*}
    \text{Revenue}    & \sim \text{Gamma}(N_{\text{active}}, \lambda)                                             \\
    \log(\lambda) = ( & \text{intercept}                                                                          \\
                      & + \beta_{\text{cohort age}} \times \text{cohort age}                                      \\
                      & + \beta_{\text{age}} \times \text{age}                                                    \\
                      & + \beta_{\text{cohort age} \times \text{age}} \times \text{cohort age} \times \text{age}) \\
    N_{\text{active}} & \sim \text{Binomial}(N_{\text{total}}, p)                                                 \\
    \textrm{logit}(p) & = \text{NN}(\text{cohort age}, \text{age}, \text{month})
\end{align*}

where $\text{NN}$ represents a neural network. Even a simple architecture with one hidden layer containing just 4 units and
sigmoid activation functions can capture the complex patterns in retention data effectively.

\subsection{Advantages of the Neural Network Approach}

This neural network approach offers several advantages:

\begin{enumerate}
    \item \textbf{Computational efficiency}: Inference can be performed using stochastic variational inference (SVI), which
          is significantly faster than the MCMC sampling required for BART models. This enables rapid model iteration and
          scaling to larger datasets.

    \item \textbf{Flexibility in inference methods}: Beyond SVI, the NumPyro framework allows for various sampling methods,
          including NUTS (No U-Turn Sampler) for full Bayesian inference when needed, as well as integration with other
          JAX-based probabilistic programming tools like BlackJax (\cite{cabezas2024blackjax}). To be fair, this can
          also be done with PyMC thanks to the PyTensor backend.

    \item \textbf{Comparable predictive performance}: Experiments on the same synthetic dataset show that the neural network
          approach produces similar retention and revenue predictions as the BART-based model, with well-calibrated $94\%$ HDIs
          that appropriately capture uncertainty.

    \item \textbf{Development workflow}: The computational efficiency enables an iterative workflow where initial model
          development and testing can use fast SVI methods, with final inference performed using full MCMC sampling if desired.
\end{enumerate}

\subsection{Limitations of Neural Networks Compared to BART}

Despite these advantages, the neural network approach does have some limitations when compared to BART:

\begin{enumerate}
    \item \textbf{Reduced interpretability}: Unlike BART, neural networks do not naturally provide partial dependence plots
          (PDP) or individual conditional expectation (ICE) plots. These visualizations, which help understand how individual
          predictors affect the target variable, require additional custom implementation with neural networks.

    \item \textbf{Architecture selection}: Neural networks require specification of the network architecture (number of
          layers, units per layer, activation functions), which introduces additional hyperparameters that must be selected,
          whereas BART requires fewer tuning decisions.
\end{enumerate}

\subsection{Practical Considerations}

The choice between BART and neural network approaches depends on the specific needs of the application:

\begin{itemize}
    \item For applications where interpretability is paramount and computational efficiency is less critical, BART may be
          preferred.

    \item For large-scale applications where inference speed is essential or when rapid model iteration is needed, the neural
          network approach with SVI offers significant advantages.

    \item In some cases, a hybrid approach might be valuable—using the faster neural network model for initial exploration
          and prototyping, then moving to BART for final analysis when interpretability is needed.
\end{itemize}

The implementation details and complete code examples for the neural network approach can be found in
\cite{orduz_revenue_retention_numpyro}.

\section{Conclusion and Future Directions}

The ability to accurately forecast retention and revenue metrics represents a significant competitive advantage in today's
business environment. In this paper, we have presented a novel Bayesian approach to modeling cohort-level retention and
revenue that addresses many of the limitations inherent in traditional methodologies. By combining the flexibility of
Bayesian additive regression trees with the interpretability of linear models, our approach offers both analytical power and
practical utility. Our framework provides several distinctive advantages that merit highlighting:

\begin{enumerate}
    \item \textbf{Adaptive complexity}: The BART component automatically adjusts its complexity to match the underlying
          patterns in the retention data, capturing non-linear relationships and interactions that would be difficult to
          specify manually. Meanwhile, the linear component for revenue provides clear interpretability of key drivers,
          offering the best of both worlds—sophisticated modeling where needed and transparency where possible.

    \item \textbf{Principled uncertainty quantification}: Unlike deterministic approaches that provide only point estimates,
          our Bayesian framework generates complete posterior distributions for all quantities of interest. This allows
          decision-makers to understand the full range of potential outcomes through $94\%$ highest density intervals (HDI) and
          tailor their strategies to their risk preferences. For instance, a risk-averse business might base resource
          allocation decisions on lower quantiles of the revenue prediction distribution rather than mean estimates.

    \item \textbf{Knowledge transfer across cohorts}: The model's structure enables effective information sharing between
          cohorts, leveraging patterns from data-rich older cohorts to improve predictions for newer cohorts with limited
          history. This is particularly valuable in fast-growing businesses where the latest cohorts often represent
          significant portions of the customer base yet have the least historical data.

    \item \textbf{Customizable architecture}: The modular design allows for straightforward extensions to incorporate
          business-specific factors and external variables. Whether integrating marketing channel information, product usage
          metrics, or macroeconomic indicators, the model can adapt to diverse business contexts without fundamental redesign.
\end{enumerate}

Our experiments with synthetic data demonstrate the model's effectiveness, but the real value of this approach emerges in
practical business applications. By providing both accurate forecasts and well-calibrated uncertainty estimates through
highest density intervals (HDI), this methodology enables more informed decision-making across multiple business functions:

\begin{itemize}
    \item \textbf{Marketing teams} can optimize acquisition spending based on expected customer lifetime value, potentially
          varying their strategies seasonally based on predicted retention patterns.

    \item \textbf{Product teams} can prioritize features that target high-value cohorts or address specific drop-off points
          in the customer lifecycle.

    \item \textbf{Financial planning} becomes more robust with probabilistic forecasts that account for the inherent
          uncertainty in future customer behavior.

    \item \textbf{Customer success initiatives} can be tailored to specific cohorts based on their predicted retention
          trajectories, potentially intervening at critical points to improve outcomes.
\end{itemize}

Despite these advantages, we acknowledge some limitations that present opportunities for future research. First, while our
top-down approach efficiently models cohort-level patterns, it cannot provide individual-level predictions or personalized
insights. Businesses requiring customer-specific forecasts would need to complement this approach with individual-level
models. Second, the current framework assumes that cohort behavior patterns remain relatively stable over time, with seasonal
variations occurring around consistent trends. In rapidly evolving markets or during significant disruptions, this assumption
may not hold. Future work could explore regime-switching models or online learning approaches that adapt more quickly to
fundamental shifts in customer behavior. Third, our model currently treats cohorts as distinct entities defined solely
by their start date. An interesting extension would be incorporating cohort formation factors—such as acquisition channel,
initial product selection, or demographic characteristics—directly into the model structure, potentially uncovering more
nuanced retention and revenue patterns.

Looking ahead, several promising research directions emerge:

\begin{enumerate}
    \item \textbf{Causal modeling extensions}: Incorporating causal inference techniques to estimate the impact of
          interventions on retention and revenue would enhance the model's utility for decision support.

    \item \textbf{Multi-product ecosystems}: Extending the framework to handle customers who engage with multiple products or
          services, capturing cross-product effects on retention and spending.

    \item \textbf{Hierarchical structures}: Implementing full hierarchical Bayesian models to more formally represent the
          relationships between cohorts and potentially incorporate prior business knowledge. For instance, we can have
          retention and revenue matrices per country and we would like to model them all together through a hierarchical
          structure on both components.
\end{enumerate}

The methodology presented in this paper represents a significant step forward in cohort-based retention and revenue modeling.
By embracing the complexity inherent in customer behavior while maintaining analytical tractability, our approach bridges the
gap between sophisticated statistical techniques and practical business applications. As companies continue to recognize the
strategic importance of customer retention and lifetime value, flexible and accessible modeling approaches like the one presented
here will become increasingly essential tools in the modern business analytics toolkit.

\bibliographystyle{acm}
\bibliography{references}

\end{document}